\documentclass[a4paper]{jpconf}
\usepackage{graphicx}
\usepackage{citesort}

\begin{document}
\title{Deep inelastic scattering from string/gauge duality with soft IR cutoff and exponentially
small Bjorken parameter}

\author{E F Capossoli$^{1,2}$ and H Boschi-Filho$^2$}


\address{$^1$Departamento de F\'isica and Mestrado Profissional em Pr\'aticas da Educa\c c\~ao B\'asica (MPPEB),
Col\'egio Pedro II, 20.921-903 - Rio de Janeiro - RJ - Brazil}

\address{$^2$Instituto de F\'isica, Universidade Federal do Rio de Janeiro, 21.941-972 - Rio de Janeiro - RJ - Brazil}

\ead{eduardo\_capossoli@cp2.g12.br}

\ead{boschi@if.ufrj.br}

\begin{abstract}
In this work we use the string/gauge duality within the Softwall Model (SW). In this model a dilaton field is
introduced in the action for the fields playing the role of a soft infrared (IR) cutoff. The SW model is very useful
as it provides linear Regge trajectories for mesons. Here, using a $10-$dimensional SW model, we calculate
the corresponding structure functions for deep inelastic scattering (DIS) in which electrons are scattered off
hadrons in a kinematical regime where the hadrons are broken apart, with high virtuality $q$, in the exponentially
small $x$ (Bjorken parameter) regime. Our results for this regime are consistent with those achieved using other
holographic and non-holographic approaches.
\end{abstract}

\section{Introduction}

Deep Inelastic Scattering (DIS) has an important role in high energy experimental physics since using it we could access the internal structure of protons or other hadrons.

The DIS can be described by a scattering between a lepton $\ell$ and a target hadron with momentum $P$ producing a diversity of other hadrons. During this scattering a virtual photon of momentum $q$ is exchanged.

The relevant parameters for DIS are the virtuality of the photon, given by $q^2  = q_\mu q_\nu \eta^{\mu \nu} = -Q^2$, with metric signature $\eta^{\mu \nu} = (-, +, +, +)$, the mass $M$ of the initial hadron, such that $M^2 = - P^2$, the squared center-of-mass energy $s= -P^2_X = - (P + q)^2$ and the Bjorken variable $x$, defined by $x \equiv -q^2 /2.P.Q$.

For our purposes, the differential cross section for DIS is given by $d\sigma \propto  (\alpha^2/ q^4) L^{\mu\nu} W_{\mu \nu},$
%
\noindent where  $\alpha$ is the fine structure constant, $L^{\mu\nu} $ is the leptonic tensor, and $W_{\mu \nu}$ is the hadronic tensor, which is the quantity of interest here. In the case where the initial hadronic state is not polarized, the hadronic tensor can be written as \cite{Manohar:1992tz}:
\begin{equation}\label{f3}
W^{\mu \nu} = i \int d^4 y~ e^{i q.y} \langle P, {\cal Q}|[J^{\mu}(y), J^{\nu}(0)]|P, {\cal Q} \rangle
\end{equation}

\noindent where $|P, {\cal Q} \rangle$ represents a normalizable hadronic state with 4-momentum $P^{\mu}$  and electric charge ${\cal Q}$ of the initial hadron and $J^{\mu}$ is the electromagnetic hadronic current.

Using gauge invariance, which implies $q_{\mu}W^{\mu \nu} = 0$, together with the Lorentz covariance, the hadronic tensor in (\ref{f3}) can be decomposed as in the following:
\begin{equation}\label{f4}
W^{\mu \nu} = W_{1} \left( \eta^{\mu \nu} - \frac{q^{\mu} q^{\nu}}{q^2} \right) + \frac{2x}{q^2} W_2 \left( P^{\mu} + \frac{q^{\mu}}{2x} \right)\left( P^{\nu} + \frac{q^{\nu}}{2x} \right),
\end{equation}
\noindent where $W_1 (x, q^2)$ e  $W_2 (x, q^2)$ are the unpolarized structure functions, which describe the quark distribution of momenta inside the hadrons.

For a particular combination between $x$ and $q^2$, one can see an approximate relation between these functions called the Callan-Gross relation, so that:
\begin{equation}\label{f5}
W_2 (x, q^2) \approx 2 x \;W_1 (x, q^2).
\end{equation}
 
The canonical approach to deal with strong interactions is based on QCD. However QCD fails in the low energy limit when $ g_ {YM}> 1$. In this context we have to choose another approach to overcome this difficulty.

The $AdS/CFT$ correspondence \cite{Maldacena:1997re}, also known as string/gauge duality, proposed by Juan Maldacena, brought new ways to study strong interactions where QCD cannot be treated perturbatively.

This correspondence or duality relates a conformal Super Yang-Mills (SYM) theory with extended supersymmetry ${\cal N} = 4$ and the symmetry group $SU(N)$ for $N \to \infty$ (large $N$) in a flat Minkowski space-time in 3 + 1 dimensions with a type IIB superstring theory in a $10-$dimensional curved space, which is a five dimensional anti de Sitter space times a five dimensional hypersphere, or simply, $AdS_5 \times S^5$.

Due to the conformal invariance one cannot use the correspondence directly. In order to break this invariance one can use, for instance, two fruitful bottom-up approaches known as the hardwall and softwall models. In the first one, a hard cutoff is introduced in the $AdS$ space and a slice of this space in the region $0 \leq z \leq z_{\rm max}$ is considered, with a boundary condition at $z=z_{\rm max}$. For more information one can see \cite{Polchinski:2001tt,BoschiFilho:2002vd,BoschiFilho:2002ta,BoschiFilho:2005yh,Capossoli:2013kb,Rodrigues:2016cdb}. In the context of DIS, using the hardwall, one can see \cite{Polchinski:2002jw,Brower:2006ea,Hatta:2007he,BallonBayona:2007rs,Gao:2009ze,Brower:2010wf}.

In the second bottom-up approach, or the softwall model, a soft infrared cutoff is introduced in the action. This is done by using a decreasing exponential related to the dilatonic field. For more information one can see \cite{Karch:2006pv,Colangelo:2007pt,Li:2013oda,Capossoli:2015ywa,Capossoli:2016kcr,Capossoli:2016ydo,FolcoCapossoli:2016ejd}. Other studies for DIS have been discussed within the softwall model, finding results consistent with the literature \cite{BallonBayona:2007qr,Braga:2011wa}. For other DIS studies using holographic models see for instance \cite{BallonBayona:2010ae,Cornalba:2008sp,Pire:2008zf,Albacete:2008ze,BallonBayona:2008zi,Yoshida:2009dw,Hatta:2009ra,Avsar:2009xf,Cornalba:2009ax,Bayona:2009qe,Cornalba:2010vk,Koile:2013hba,Koile:2014vca,Koile:2015qsa}. 

Here, in this work, we discuss the exponentially small $x$ regime within the softwall model and calculate the corresponding structure functions. More details related to these calculations can be seen in \cite{Capossoli:2015sfa}. 

\section{The DIS within the Softwall Model}

Let us start this section describing the DIS within the softwall model \cite{BallonBayona:2007qr,Capossoli:2015sfa}.  In this approach the action for the fields in the $AdS_5 \times S^5$ is given by:
\begin{equation}\label{acao_soft}
S = \int d^{10} x \sqrt{-g} \; e^{-\phi(z)} {\cal L}
\end{equation}
\noindent where ${\cal L}$ is the Lagrangean density, $\phi = k z^2$ is a scalar field related to the dilaton, $g$ is the determinant of the metric $g_{MN}$ of the $AdS_5 \times S^5$ space:
\begin{equation}\label{gs}
ds^2 = g_{MN} dx^M dx^N= \frac{R^2}{z^2}(dz^2 + \eta_{\mu \nu}dy^\mu dy^\nu) + R^2 d\Omega^2_5, 
\end{equation}

\noindent $z$ is the holographic coordinate, $d\Omega^2_5$ is the angular measure on $S^5$, and $R$ is the $AdS_5$ space radius.

In order to deal with the DIS within the softwall model it is convenient to split the action Eq.(\ref{acao_soft}) in different parts. The first part, with 5-dimensional gauge fields $A_m = (A_{\mu}, A_z )$, which does not depend on the coordinates of $S^5$ space, is given by:
\begin{equation}\label{f15}
S = -  \int d^{10} x \sqrt{-g} \; e^{-\phi(z)} \frac{1}{4} F^{m n} F_{m n}.
\end{equation}

The second one, with scalar fields $\Phi$, which describe the initial and final hadrons, which for simplicity are considered both as spinless, is given by:
\begin{equation}\label{f26}
S = \int d^{10} x \sqrt{-g} \; e^{-\phi(z)} ( \partial_m \Phi \partial^m \Phi + m^2_5 \Phi^2).
\end{equation}

From the EOM of the action Eq. (\ref{f15}), with a convenient choice of a Lorentz-like gauge \cite{BallonBayona:2007qr}, one gets:
\begin{equation}\label{f22}
A_{\mu}(y^{\mu},z) =  \eta_{\mu}\; k \; \Gamma(1 + \frac{q^2}{4k}) \;e^{iq.y} \; z^2 \; {\cal U}(1 + \frac{q^2}{4k}; 2; kz^2)\;,
\end{equation}
\begin{equation}\label{f23}
A_z (y^{\mu},z) =  \frac{i q. \eta}{2} \; \Gamma(1 + \frac{q^2}{4k}) \;e^{iq.y} \; z \; {\cal U}(1 + \frac{q^2}{4k}; 1; kz^2).
\end{equation}
\noindent where $\Gamma(x)$ is the Gamma function and ${\cal U}(a; b; c)$ is the Tricomi confluent hypergeometric function.

From the EOM of the action Eq. (\ref{f26}), with the ansatz, $\Phi (y_{\mu},~ z,~ \Omega) = e^{iP.y}\psi(z, \Omega),$ one gets for the initial state of the hadron:
\begin{equation}\label{f33}
\Phi_i (y_{\mu},~ z,~ \Omega) = \left[\frac{2k^{\Delta - 1}}{\Gamma( \Delta - 1)}\right]^{1/2} \frac{1}{R^4}\; e^{iP.y}\; z^{\Delta} \; \psi(\Omega).
\end{equation}
For the final state, one has:
\begin{equation}\label{f35}
\Phi_X (y_{\mu},~ z,~ \Omega) = \left[\frac{2k^{\Delta - 1} \Gamma(\frac{s}{4k} - \frac{\Delta}{2} + 1)}{\Gamma( \frac{s}{4k} + \frac{\Delta}{2} - 1)}\right]^{1/2} \frac{1}{R^4}\; e^{iP_X.y}\;z^{\Delta} \; L^{\Delta - 2}_{n_X}(kz^2)\;\psi(\Omega), 
\end{equation}
\noindent where $\Delta$ is the conformal dimension of an operator associated with the initial and final hadrons \cite{BallonBayona:2007qr} and $L^m_n(y)$ are the associated Laguerre functions.

\section{The exponentially small$-x$ regime in the Softwall Model }

The DIS in the exponentially small $x$ regime is characterized by multiple pomeron exchange represented by gravitons in the AdS/CFT correspondence \cite{Polchinski:2002jw}.

The dominant contribution at high energies to the string scattering amplitude in the $10-$dimensional SW model is given by:
\begin{equation}\label{f38}
S_{{\rm string}} = \int d^{10}x \sqrt{-g} \; e^{-\phi(z)} {\cal L}_{{\rm eff,string}}
\end{equation}

\noindent which is identified with the amplitude of the forward Compton scattering in four dimensions and can be written as:
\begin{equation}\label{f39}
\eta_{\mu} \eta_{\nu} T^{\mu \nu} (2 \pi)^4 \delta^4 (q - q') = S_{{\rm string}},
\end{equation}
\noindent where $T^{\mu \nu}$ is a tensor which has the same decomposition of the hadronic tensor $W^{\mu \nu}$ presented in Eq.(\ref{f4}) and 
$S_{{\rm string}}$ is in the following:
\begin{eqnarray}\label{f41}
S_{{\rm string}} & = & \frac{1}{8} \int d^{10}x \sqrt{-g} \; e^{-\phi(z)} \Bigl\{ 4 v^a v_a \partial_m \Phi F^{mn} F_{pn} \partial^p \Phi \nonumber \\
                 & & - \left( \partial^M \phi \partial_M \Phi v^a v_a + 2 v^a \partial_a \Phi v^b \partial_b \Phi \right) F^{mn} F_{mn} \Bigl\} G|_{t=0}
\end{eqnarray}
\noindent where $v^a$ are the Killing vectors of the compact  $S^5$ space (or more generically $W$), $F^{mn}$ is associated with an incoming photon with $4-$momentum $q_{\mu}$ and an outgoing one with $4-$momentum $q'_{\mu}$ and $\Phi$ represents the incoming and outgoing scalars state with $4-$momentum $P^{\mu}$ and $P^{\mu}_X$, respectively. 

After some manipulations, the imaginary part of $S_{{\rm string}}$ takes the form:
\begin{eqnarray}\label{f76}
{\rm Im } \;S_{{\rm string}} &  = & (2 \pi)^4 \delta (q - q') \;\eta_{\mu} \eta{\nu} \;\frac{\pi \rho R^{8}(x)^{{\alpha' |\xi|}/2}}{4s} \; a^2 \;  \Gamma^2 \left(1 + \frac{q^2}{4k}\right) \nonumber \\
                             &    & \times \;\Biggl\{ \displaystyle\sum_{m=1}^{\infty} e^{-kz^2_m}\;  \frac{z^{2 \Delta + 2}_m}{4} \; {\cal U}^2 \left(1 + \frac{q^2}{4k}; 1; kz^2_m \right) (-q^2)  \left[p^{\mu} - \frac{p.q}{q^2} q^{\mu} \right] \left[p^{\nu} - \frac{p.q}{q^2} q^{\nu} \right]  \nonumber \\
                             &    & + \; k^2 \; \displaystyle\sum_{m=1}^{\infty}\;  e^{-kz^2_m}\; z^{2 \Delta + 4}_m\; {\cal U}^2 \left(1 + \frac{q^2}{4k}; 2; kz^2_m \right) (p.q)^2 \nonumber \\
                             &    & \times \;\left[\eta^{\mu \nu} - \frac{(p^{\nu}q^{\nu} +  p^{\nu} q^{\mu})}{p.q} + \frac{q^2}{(p.q)^2} p^{\mu} p^{\nu} \right] \Biggl\}.
\end{eqnarray}
Comparing with Eq. (\ref{f39}), one has:
\begin{eqnarray}\label{f81}
{\rm Im } \;T^{\mu \nu} &  = & \frac{\pi \rho R^{8}(x)^{{\alpha' |\xi|}/2}}{4s} \; a^2 \; \frac{q^4}{4 x^2} \Biggl\{ \left[\eta^{\mu \nu} - \frac{q^{\mu}q^{\nu}}{q^2}  \right] \;{\cal I}_2 \nonumber \\
                        &    & + \; \left[p^{\mu} + \frac{q^{\mu}}{2x} \right] \left[p^{\nu} + \frac{q^{\nu}}{2x} \right] 4x^2  \left( {\cal I}_1 + \frac{{\cal I}_2}{q^2} \right) \Biggl\} 
\end{eqnarray}

\noindent where
\begin{eqnarray}\label{f78}
{\cal I}_1 &  \equiv & \frac{1}{4} \; \Gamma^2 (j) \;\displaystyle\sum_{m=1}^{\infty} e^{-kz^2_m}\;  z^{2 \Delta + 2}_m \; {\cal U}^2 \left(j ; 1; kz^2_m \right) \\
{\cal I}_2 &  \equiv & k^2 \; \Gamma^2 (j) \; \displaystyle\sum_{m=1}^{\infty}\;  e^{-kz^2_m}\; z^{2 \Delta + 4}_m\; {\cal U}^2 \left(j ; 2; kz^2_m \right) \label{f666}
\end{eqnarray}
\noindent Consequently, one can write the structure functions for DIS within the softwall model in the exponentially small $x$ regime, so that:
\begin{eqnarray}\label{f82}
W_1 (x, q^2)&  = & \frac{\pi^2 \rho k^{\Delta - 1}}{4 \Gamma (\Delta - 1)} \; \frac{q^4 (x)^{{\alpha' |\xi|}/2}}{s x^2} \;{\cal I}_2 \nonumber \\
W_2 (x, q^2)&  = & \frac{\pi^2 \rho k^{\Delta - 1}}{4 \Gamma (\Delta - 1)} \; \frac{q^4 (x)^{{\alpha' |\xi|}/2}}{s x^2}  \;(2 x q^2) \left( {\cal I}_1 + \frac{{\cal I}_2}{q^2} \right).
\end{eqnarray}
In order to compare our results with ref.\cite{Polchinski:2002jw} one can use Hypergeometric functions' properties and consider the approximations in which $x$ is exponentially small and $q^2$ is large \cite{Capossoli:2015sfa}, so that one can rewrite Eqs.(\ref{f82}) as:
\begin{eqnarray}\label{f112}
W_1 (x, q^2)  & \approx & \frac{\pi^2 \rho \; (x)^{{-2 + \alpha' |\xi|}/2}}{8 \;(4 \pi g_s N)^{1/2}\; \Gamma (\Delta - 1)}  \left(\frac{k}{q^2}\right)^{\Delta - 1} \; {\cal I}_{1,\;2\Delta + 3} 
\\
W_2 (x, q^2) & \approx & \frac{\pi^2 \rho \; (x)^{{-1 + \alpha' |\xi|}/2}}{4 \;(4 \pi g_s N)^{1/2}\; \Gamma (\Delta - 1)} \left(\frac{k}{q^2}\right)^{\Delta - 1} \;\left({\cal I}_{0,\;2\Delta + 3} + {\cal I}_{1,\;2\Delta + 3}  \right),\label{f115}
\end{eqnarray}
\noindent where 
\begin{equation}\label{f109}
{\cal I}_{r,s} \equiv \int_{0}^{\infty} dw \;w^s \; K^2_r(w) = 2^{(s - 2)} \; \frac{\Gamma(\frac{s+1}{2} + r)\; \Gamma(\frac{s+1}{2} - r) \; \Gamma^2(\frac{s+1}{2})}{\Gamma (s+1)}.
\end{equation}
Computing the ratio of these structure functions one finds:
\begin{equation}\label{f118}
\frac{ W_2 (x, q^2)}{W_1 (x, q^2)}   \approx  2x \; \left( \frac{2 \Delta +3 }{\Delta + 2} \right).
\end{equation}

Finally, one can see that this ratio is in agreement with the one found in \cite{Polchinski:2002jw} within the hardwall model.

\section{Last comments}

In this work we have used the AdS/CFT correspondence to study DIS with an exponentially small Bjorken parameter within the Softwall model. The correspondence proved itself to be an excellent tool to tackle QCD out of the perturbative regime. Furthermore we reproduced the results found within the hardwall model.

The exponentially small $x$ regime is also important to study the QCD phase diagram. This problem was discussed within the hardwall model in \cite{Hatta:2007he}. Since the hardwall and softwall models have different spectra, in particular leading to different Regge trajectories, we studied the saturation line in the softwall model to see if it leads to any different behavior. The complete details and references can be found in \cite{Capossoli:2015sfa}. 

For recent discussions on holographic approaches to DIS, see for instance \cite{Jorrin:2016rbx,Kovensky:2016ryy,Kovensky:2017oqs,Amorim:2018yod}. 

\section{Acknowledgments}
H. B.-F. is partially supported by Conselho Nacional de Desenvolvimento Cient\'\i fico e Tecnol\'ogico (CNPq) and Coordena\c c\~ao de Aperfei\c coamento de Pessoal de N\' \i vel Superior (Capes) (Brazilian Agencies). 

\section*{References}
\bibliography{iopart-num}

\end{document}